\documentstyle[prb,aps,epsf]{revtex}   
\begin{document}
\twocolumn[\hsize\textwidth\columnwidth\hsize\csname %
@twocolumnfalse\endcsname

\title{Interchain Coupling Effects and Solitons in CuGeO$_3$}
\author{Jun Zang$^{1}$, Sudip Chakravarty$^{2}$, A.R. Bishop$^{1}$}
\address{$^{1}$
    Theoretical Division and Center for Nonlinear Studies, MS B262\\
    Los Alamos National Laboratory, 
    Los Alamos, NM 87545}
\address{$^{2}$
Department of Physics and Astronomy, University of California Los
Angeles
\\ Los Angeles, CA 90095-1547}
\date{\today}
\maketitle
\draft
\begin{abstract}
The effects of interchain coupling  on
solitons and soliton lattice structures
in  CuGeO$_3$ are explored. It is shown that
interchain coupling  substantially increases
the soliton width and changes the soliton lattice structures
in the incommensurate phase.  It is proposed that the
experimentally observed large soliton width in  CuGeO$_3$ is mainly due to
interchain coupling effects. 
\end{abstract}

\phantom{.}

]

\narrowtext
\pagebreak

The  inorganic  spin-Peierls (SP)
system, CuGeO$_3$, has attracted much attention\cite{sp-cugeo3} recently.
Pure
CuGeO$_3$  has a SP transition at $T_{\rm sp}\simeq 14.3K$.
Below  $T_{\rm sp}$, 
the system is in a dimerized spin singlet state, and the
gap to spin triplet excitations is $\Delta \sim 24.5K$ \cite{js,delta}.
It has been argued\cite{dobry,sudip} that the spin susceptibility 
measurements in
this system  can be understood only if the nearest neighbor Heisenberg
Hamiltonian is augmented by a substantial next-nearest-neighbor (NNN) coupling,
$J_2\sim 0.2J$, where $J$ is the nearest neighbor coupling. The NNN coupling has
been shown  to be important in  other experiments \cite{j2-exp} as well. 
Recently, the soliton lattice structure in CuGeO$_3$ was
studied  using neutron 
and NMR measurements \cite{sol-l,nmr}.
An important observation is that the soliton width $\xi\sim 13.6a$, where $a$
is the lattice spacing, is much larger than the theoretical prediction
\cite{fuku-ham} for the unfrustrated SP chain, $\xi_0=Ja\pi/(2\Delta)\sim 8a$.

The soliton width in CuGeO$_3$ can be changed by interchain coupling and
frustration $J_2$. Without interchain coupling, the soliton width
is generally $\xi \propto {\bar v}/\Delta$, where $\bar{v}$ is the
spin wave velocity and $\Delta$ is the {\em triplet} gap in the SP chain.
From nonlinear simga model\cite{affleck}, 
${\bar v}={\bar v}_{0}\sqrt{1-4J_2/J_1}$, and the
dimensionless coupling constant \cite{note-sigma} 
$g_s(J_2)=g_s(0)/\sqrt{1-4J_2/J_1}$. So, the frustration
will decrease the soliton width.
One can show that the soliton width $\xi=Ja{\bar v}/\Delta$.
The triplet gap $\Delta$ will depend on both SP coupling and
frustration. However, to compare
with experiments, the triplet gap $\Delta\sim 24.5K$, while $\bar{v}$
decreases with increase of frustration, so the frustration
will make the soliton width in a CuGeO$_3$ chain smaller than
$\sim 8a$. Thus, the observed
soliton width $\xi/\xi_0 \sim 1.7$ is not an effect 
due to the frustration $J_2$ in CuGeO$_3$.

It is 
known that the interchain coupling 
in CuGeO$_3$ is not small \cite{js}: 
$J_c\simeq 120K$, $J_b\simeq 0.1J_c$, and $J_a\simeq -0.01J_c$.
The phase diagram of the unfrustrated
SP system in the presence of interchain coupling was studied by Inagaki and
Fukuyama\cite{fuku-inter}. In the present paper we calculate
the effect of interchain coupling on soliton width. Although the singlet SP
ground state is destroyed by large interchain coupling, in the strongly
dimerized regime weak interchain coupling cannot affect the spin singlet
ground state.  Similarly, weak interchain coupling has very
little effect on the susceptibility above $T_{\rm sp}$. From mean field
theory, it can be shown that the effects are typically of the order of 3\%.

The low energy excitations involving 
broken spin singlets are more sensitive to interchain coupling. Once a singlet
pair is broken in one chain, it is  energetically less costly to break another
pair in a neighboring chain due to the interchain coupling. 
Thus, the interchain
coupling will affect the low energy excitations in the dimerized phase.
For the same reason, the interchain coupling
can lead to an increased soliton width: the local magnetization
of a soliton in SP chain is 
$M(x)\sim \cosh^{-1}(x/\xi)$ ({\em cf.} below), and the gain in the interchain
coupling energy increases with the increased soliton width $\xi$, hence 
an increased soliton width.

In this paper, the  effects due to the interchain coupling
are studied  using bosonization.
We consider the Hamiltonian:
\begin{eqnarray}
{\cal H}&=&\sum_{i,j} \left[ J(1+\beta u_{i,j})\vec{S}_{i,j}
\cdot\vec{S}_{i+1,j} +\alpha_0 J\vec{S}_{i,j}
\cdot\vec{S}_{i+2,j}\right]
\nonumber \\
&+& \gamma J\sum_{i,j,\mu} \vec{S}_{i,j}\cdot\vec{S}_{i,j+\mu}
+{K\over2}\sum_{i,j}u_{i,j}^2,
\end{eqnarray}
where $\vec{S}_{i,j}$ ($u_{i,j}$) is the spin operator (lattice distortion)
of site-$i$
in chain-$j$, and $K$ is the spring constant. 
$\beta$ is the spin-lattice coupling constant.
$\alpha_0=J_2/J$ is the strength of frustration and $\gamma=J_{\perp}/J$
is the interchain coupling constant. 
Note that we have used the adiabatic approximation
for the phonons.

The interchain coupling is treated in the mean-field
approximation\cite{fuku-inter,chain-mf}:
$\vec{S}_{i,j}\cdot\vec{S}_{i,j+\mu}=S^z_{i,j}
\langle S^z_{i,j+\mu}\rangle$.
Thus, the Hamiltonian is transformed into a single chain model
and the interchain coupling term transforms into
$
-\sum_i\gamma J\tilde{Z}h_iS_i^z+{1\over2}\gamma J Z\sum_i h_i^2
$,
where
$
h_i=\langle S^z_{i,j+\mu}\rangle = -(\tilde{Z}/Z)
\langle S^z_{i,j}\rangle
$
is the effective field. We have introduced the notion of an effective
coordination number $\tilde{Z}$ to account for 
$\langle S^z_{i,j+\mu} \rangle \neq
-\langle S_{i,j}^z\rangle$. In the soliton lattice phase, $\tilde{Z}$
is site dependent.
The effective Hamiltonian
is  then transformed to a fermionic Hamiltonian
using the Jordan-Wigner transformation, and the fermionic model is  bosonized.
The resulting continuum Hamiltonian is \cite{jun}:
\begin{eqnarray}
{{\cal H}\over Ja} &=& {\bar{v}\over 2\pi K_{\rho}}
\int dx [K^2_{\rho} (\partial_{0}\phi)^2+(\partial_x\phi)^2]
+\int dx {Ku^2(x)\over2Ja^2}
\nonumber \\
&-&{g_3\over2a^2} \int dx \cos(4\phi) 
-{\beta\over a^2}\int dx u(x)\sin(2\phi)
\nonumber \\
&-& {\gamma\tilde{Z}\over a^2} \int dx h(x)\cos(2\phi)
+ \int dx{\gamma Z\over2a^2} h^2(x),
\label{eq:ham1}
\end{eqnarray}
where $a$ is the short distance cutoff, $u(x)=(-1)^i u_i/a$ and 
$h(x)=(-1)^i h_i/a$ 
are slowly varying variables. 
The bare value of the Umklapp term
is $g_3=1-3\alpha_0$ and  we shall assume that the
renormalized $g_3$ has approximately the same $\alpha_0$-dependence.

Here $\bar{v}$ is the spin wave velocity
and $K_{\rho}$ is a critical exponent.
For the unfrustrated AFM Heisenberg spin
chain,  the following renormalized values should be used \cite{cross}:
$\bar{v}=\pi/2$ and $K_{\rho}=1/2$. 
However, the values of $\bar{v}$ and $K_{\rho}$ in the frustrated
Heisenberg model is dependent on $\alpha_0$.
With increase of frustration, the spin wave velocity 
decreases. In nonlinear sigma model \cite{affleck}, 
$\bar{v}=v_0\sqrt{1-4\alpha_0}$ for small $\alpha_0$.
In a recent numerical calculation,
it is shown that the variation of {\it renormalized} $\bar{v}$ with $\alpha_0$
 is approximately \cite{gros}
$\rho_{v}=2\bar{v}/\pi\simeq (1-1.12\alpha_0)$.
Although we don't know $K_{\rho}$ at finite $\alpha_0$ precisely,
it is known \cite{haldane} that $K_{\rho} \geq 1/2$ at
small frustration $\alpha_0<\alpha_0^c\sim 0.3$.
Define $\rho_K=2K_{\rho}-1$,
then $\rho_K\ll 1$ and $\rho_v \alt 1$.

For the ground state, $u(x)=u_0$, $h(x)=h_0$, and $\tilde{Z}=Z$.
If there is no Umklapp interaction $g_3$, the model (\ref{eq:ham1})
is exactly soluble by Bethe Ansatz\cite{exact}. Due to the
$g_3$-term, an approximation has to be employed. We
use the self-consistent Gaussian approximation, which was
previously shown to be reliable\cite{fuku-ham} for the SP chains.  
To this end, we write
$\phi(x)=\phi_s(x)+\tilde{\phi}(x)$, where $\phi_s$ is the semiclassical solution
and $\tilde{\phi}$ is the fluctuation around $\phi_s$. We retain
$\tilde{\phi}$ to quadratic order in the action.
The first order term vanishes at the saddle point. Then,
\begin{equation}
\partial^2_x(4\phi_s)+{1\over\xi^2}\sin(4\phi_s)=0
\label{eq:sad}
\end{equation}
where
\begin{equation}
{1\over\xi^2}={1\over\xi^2_0}-{4\gamma\tilde{Z}^2\over a^2Z}\sigma^2,
\end{equation}
and
\begin{equation}
{1\over\xi^2_0 }={4\beta^2J\over a^2K}\sigma^2-{8g_3\over a^2}\sigma^4.
\end{equation}
The quantity $\sigma$ is given by 
\begin{equation}
\sigma=e^{-2\langle\tilde{\phi}^2\rangle}.  
\end{equation}
$\sigma$ describes the renormalization of $g_1$ and $g_3$ 
due to fluctuations of $\tilde{\phi}$  around $\phi_s$.
Note the derivation is appropriate for systems with a finite gap,
otherwise, $\sigma$ has {\it infrared} divergence.
In deriving these  equations,
we have used the self-consistent equations
\begin{eqnarray} 
u(x)&=&(\beta J/K)\sigma\sin(2\phi_s), \\
h(x)&=&(\tilde{Z}/Z)
\sigma\cos(2\phi_s).
\end{eqnarray}

The ``uniform'' ground state (i.e. $u_i=(-1)^iu_0a$ and $h_i=(-1)^i h_0a$)
calculated from Eq. (\ref{eq:sad})
corresponds to $\sin(4\phi_s)=0$. Therefore, either $\phi_s=0$ or $\pi/4$.
When $\phi_s=0$, $u(x)=0$, $h_0\neq0$, the system has long range
AFM order; when $\phi_s=\pi/4$, $h(x)=0$, $u_0\neq0$, the system 
is dimerized. 
The spin-lattice coupling and inter-chain coupling strengths dictate the
solution that has the lowest energy. 
It is clear that on symmetry grounds these
two homogeneous solutions are mutually exclusive.

The calculation of ground state energy and  excitation spectrum in
the self-consistent Gaussian approximation is straightforward
and can be read off from Ref.\cite{fuku-ham}, so we will just
state the results. The excitation
spectrum corresponding to uniform $\phi_s$ is 
\begin{equation}
\omega_q =Ja\bar{v}\sqrt{q^2+q^2_0}
\label{eq:del}
\end{equation}
In the dimerized phase $\phi_s=\pi/4$ and $\phi_0=0$, and
it can be easily shown
that $q_0=1/\xi_0$.
From Eq. (\ref{eq:del}),
the spin triplet excitation gap is $\Delta=Ja{\bar v}q_0$.
The scaling of the spin triplet excitation gap is 
\begin{eqnarray}
\Big({\Delta\over 2\pi J\bar{v}}\Big)^{1-\rho_K}
 \propto  \delta^{(2-2\rho_K)/(3-\rho_K)}\, ,
\label{eq:delta}
\end{eqnarray}
where $\delta=\beta u$ is the bond alternation induced by
lattice dimerization.
For the unfrustrated SP chain, $K_{\rho}=1/2$, $\rho_K=0$, and we
obtain the correct scaling relation \cite{note-log} 
$\Delta\propto \delta^{2/3}$.

Comparing the ground state energies between dimerized phase
($\phi_s=\pi/4$) and AFM phase ($\phi_s=0$), we get, for
$\rho_K \ll 1$, that the crossover from  $\phi_s=0$ 
to $\phi_s=\pi/4$ is determined by
\begin{equation}
C_K\equiv
{J\gamma Z\over\Delta\rho_v}
\Big({\Delta\over \pi^2 J\rho_v}\Big)^{\rho_K}=1.
\label{eq:c-o}
\end{equation}
If $C_K <1$ 
then $\phi_s=\pi/4$, $\delta =\beta u$, $h=0$, and the system is dimerized;
otherwise, 
$\delta=\gamma Zh$, $u=0$, and the ground state has
long range AFM order. A  similar critical value
of the interchain coupling for the unfrustrated 
SP model was derived by Inagaki and Fukuyama \cite{fuku-inter}.
Using $\Delta\simeq 24.5K$ and $J\simeq 120K$, we find that
$C_K \sim 0.98\rho_v (0.021/\rho_v)^{\rho_K}<1$,
 which is consistent with the fact
that  CuGeO$_3$ is dimerized at zero temperature. 
Since $\rho_v$ ($\rho_K$), which is less (greater) than 1(0),  decreases (increases)
with increase of frustration, the frustration pushes the SP-N\'eel
transition to stronger interchain coupling.
If $\rho_v\alt 1$ and $\rho_K\ll 1$,  the interchain coupling is close
to the critical value for the long range AFM order to be realized.
This is probably significant, as  materials with only 
3\%  Zn exhibit  long range AFM order.

Consider the nonuniform solutions of Eq. (\ref{eq:sad}),
which correspond to the solitonic excitations of the dimerized phase. In zero
magnetic field, the soliton solution has lower energy \cite{fuku-ham}
than the spin-triplet excitation with the gap $\Delta$. In this case,
we can continue to  assume that $\tilde{Z}/Z$ is a constant, and
the soliton solution from  Eq. (\ref{eq:sad}) is
\begin{equation}
u(x)\propto \sin(2\phi_s) = \pm\tanh{x\over\xi}
\label{eq:s-u}
\end{equation}
with the  half-width
\begin{equation}
\xi={Ja\bar{v}\over\Delta}/\sqrt{1-(\tilde{Z}/Z)^2C_K}\, .
\label{eq:xi}
\end{equation}
The corresponding magnetization, $M(x)$, is
\begin{equation}
M(x)=\cosh^{-1}(x/\xi)/(2\pi\xi)
\end{equation}
and the staggered magnetization, $N(x)$, is
\begin{equation}
N(x)=(-)^x\sqrt{q_0a\over2\pi}\cosh^{-1}(x/\xi).
\end{equation}
From Eq. (\ref{eq:xi}), we can see that the soliton width is
increased by interchain coupling, while the effects of
frustration $\alpha_0 =J_2/J$  appear only in $\Delta$ and ${\bar v}$.

Without interchain
coupling, the soliton width is $\xi_0=Ja\pi\rho_v/(2\Delta) \alt 8a$
for CuGeO$_3$. 
With $\tilde{Z}=Z$, 
$\langle S^z_{i,j+\mu} \rangle =
-\langle S_{i,j}^z\rangle$, 
the soliton excitation
described by Eq. (\ref{eq:s-u}) is an array of
antiferromagnetically coupled solitons in a direction perpendicular
to the SP chains.
As  discussed in the introduction, the
spins are antiferromagnetically ordered inside the soliton, 
therefore increased  soliton
size will lead to a gain in the interchain coupling energy. This is the physical
origin of the increased soliton width in the presence of interchain coupling. 
When the interchain coupling
is so strong that the soliton width diverges, i.e. 
$C_K=1$, the system crosses over
to the long range antiferromagnetically  ordered phase. This
criterion $C_K= 1$, derived from consideration of the
soliton size, is the same as that derived  from Eq. (\ref{eq:c-o}) for small
$\rho_K$.

If the solitons are excited locally without forming a regular array,
$\langle S^z_{i,j+\mu} \rangle \neq
-\langle S_{i,j}^z\rangle$, and
$\tilde{Z}/Z$ is smaller than unity. The soliton size will still be
increased with the increased
interchain coupling $\gamma$, and AFM domains will form and  increase in size.
We propose that for the solitonic excitations due to magnetic or non-magnetic
impurities similar AFM domains will form. The percolation of these AFM 
domains must be relevant to impurity induced AFM ordering
in the doped CuGeO$_3$ systems \cite{neel,neu-coex}.

\begin{figure}
\centerline{
\epsfxsize=8.0cm \epsfbox{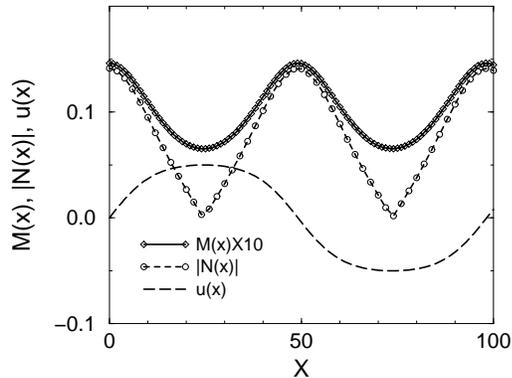}
}
\caption{The lattice distortion and magnetizations in the
soliton lattice phase with $\xi=13.6$, $k=0.8$. The scale 
of $u(x)$ is arbitrary.
\label{fig} }
\end{figure}

In a high magnetic field, there will be a transition
from the zero field commensurate dimerized phase to the incommensurate
soliton lattice phase \cite{sol-la}. 
The ground state of this soliton lattice
phase corresponds to the self-consistent periodic solutions
of the coupled nonlinear equations similar to Eq. (\ref{eq:sad}).
Here, we consider only the qualitative
aspects of the effect of the interchain coupling in the soliton lattice phase. 
To this end, we assume a constant $\xi$ in Eq. (\ref{eq:sad})
for each chain. The effect of the interchain coupling is to
renormalize $\xi$, as in Eq. (\ref{eq:xi}).
The periodic solution
of Eq. (\ref{eq:sad}) is:
\begin{eqnarray}
u(x)&\propto& sn({x\over k\xi},k),\\
M(x)&=& {1\over 2\pi k\xi} dn({x\over k\xi},k),\\
N(x)&=& (-)^x \sqrt{q_0a\over2\pi} cn({x\over k\xi},k) ,
\label{eq:sol-l}
\end{eqnarray}
where $\xi$ is the soliton width defined in Eq. (\ref{eq:xi}), and $sn$, $dn$,
and $cn$ are the Jacobi elliptic functions\cite{book-sn}. 
The inter-soliton distance is $2K(k)k\xi$ [see
Fig.(\ref{fig})], where the modulus  $k$ of the 
elliptic integral $K$ is related
to the total magnetization induced by the external magnetic field that can
be derived from the minimization of the energy \cite{sol-la}.
These equations are  the same as those of the soliton lattice
without interchain coupling\cite{note-sn}. 
Close to the
commensurate-incommensurate transition, $M(x)\sim 0\ll |N(x)|$,
we can assume that these equations still approximately
describe the soliton lattice in the presence of interchain coupling:
each chain has a magnetization $M(x)$, and the neighboring chains
have staggered magnetizations $N(x)$ and $-N(x)$.
The effective coordination number $\tilde{Z}/Z$ is smaller than unity.
Thus, $\tilde{Z}/Z={\rm constant}$ is a
reasonable approximation close to the
commensurate-incommensurate phase transition, where most
of the experiments are carried out. 

In the experiment of Kiryukhin {\em et al.}\cite{sol-l}, the value of
$\xi$ is measured in the soliton lattice phase.
The measured value of $\xi$ is $\xi=(13.6\pm0.3)a$, which is substantially
larger than the value $\xi_0\sim 8a$ without interchain coupling.
If we use an average $\tilde{Z}/Z$, from $\xi=(13.6\pm 0.3)a$
we estimate $\tilde{Z}/Z\sim 0.8$ at $\alpha_0=0$.
With finite $\alpha_0$, the estimated $\tilde{Z}/Z$ will be larger.
In these experiments the inter-soliton distance is $40-70 \rm a$,
so $k\sim 0.7-0.9$. The magnetization $M(x)$ and staggered
magnetization $N(x)$ are  shown in Fig. (\ref{fig}) for $k=0.8$. The ratio
$\langle M(x)\rangle/\sqrt{\langle N^2(x)\rangle} \sim 0.1$, which
is the right order of magnitude to give 
effective\cite{note-exp} $\tilde{Z}/Z\sim 0.8-1.0$. In the NMR experiments\cite{nmr},
the maximal value of the effective local spin is measured to be $S_{max}\sim
0.065$, which is smaller than the value corresponding to a single chain given
by  $\sqrt{\Delta/(J\pi^2\rho_v)}\agt 0.14$. However, because
the solitons in the neighboring chains are antiferromagnetically coupled,
the effective local magnetization at each site is reduced
by the interchain effects; a reduction factor of the order of $2-3$ of 
$S_{\rm max}$ is quite reasonable.

In conclusion, we have studied  the combined effects of the frustration
and interchain coupling on solitons and soliton-lattice excitations
in CuGeO$_3$.
We find that  the interchain coupling can substantially increase the soliton 
size, while the NNN frustration will decrease it. 
When the interchain coupling strength is close to the SP-N\'eel  transition, the
size of the soliton diverges. The analysis  of the soliton structure presented
here is consistent with the experimental observations\cite{sol-l,nmr}.

We thank G. Baker, S. Kivelson, and A.B. Saxena for helpful discussions.
The work at Los Alamos 
was performed under the auspices of the U. S. DOE. 
J. Z.  and S. C. thanks  the Aspen Center for
Physics where this work was initiated. S. C. acknowledges support from the
National Science Foundation, Grant No. DMR-9531575.

%\end{multicols}


\begin{thebibliography}{10}
\bibitem{sp-cugeo3}
M. Hase, I. Terasaki, and K. Uchinokura, Phys. Rev. Lett.
{\bf 70}, 3651 (1993) and
Phys. Rev. B {\bf 48}, 13 (1993) 9616.

 \bibitem{js}
 M. Nishi, O. Fujita, and J. Akimitsu, Phys. Rev. B {\bf 50}, 6508 (1994);
 L. P. Regnault {\em et al.}, Phys. Rev. B {\bf 53}, 5579 (1996).
 
 \bibitem{delta} 
 K. Hirota {\em et al.}, Phys. Rev. Lett. {\bf 73}, 736 (1994);
 O. Fujita {\em et al.}, Phys. Rev. Lett. {\bf 74}, 1677 (1994).
 
\bibitem{dobry}
J. Riera and A. Dobry, Phys. Rev. B {\bf 51} 16098 (1995).

\bibitem{sudip}
 G. Castilla, S. Chakravarty and V. J. Emery, Phys. Rev.
Lett. {\bf 75}, 1823 (1995).

 \bibitem{j2-exp}
V.N. Muthukumar {\em et al.}, Phys. Rev. B {\bf 54}, R9635 (1996);
B. Buchner {\em et al.}, Phys. Rev. Lett. {\bf 77}, 1624 (1996).

\bibitem{sol-l}
V. Kiryukhin {\em et al.}, Phys. Rev. Lett. {\bf 76}, 4608 (1996);
V. Kiryukhin {\em et al.}, Phys. Rev. B (to be published).

\bibitem{nmr} 
Y. Fagot-Revurat {\em et al.}, Phys. Rev. Lett. {\bf 77}, 1861 (1996).

\bibitem{fuku-ham}
T. Nakano and H. Fukuyama, J. Phys. Soc. Jpn {\bf 49} 1679 (1980);
{\bf 50} 2489 (1980).

\bibitem{affleck} I. Affleck, in {\it
Fields, Strings and Critical Phenomena}, eds E. Brezin and
J. Zinn-Justin (North-Holland, Amsterdam, 1989).

\bibitem{note-sigma} In nonlinear sigma model, ${\bar v}=\sqrt{\xi_{\perp}\rho_s}$,
$g_s=\sqrt{\xi_{\perp}/\rho_s}$. With finite $J_2$, 
$\xi_{\perp}\rightarrow \xi_{\perp}$, $\rho_s\rightarrow \rho_s (1-4J_2/J)$.
So the changes of $g_s$ and $\bar{v}$ have the same origin.

\bibitem{fuku-inter}
S. Inagaki and H. Fukuyama, J. Phys. Soc. Jpn {\bf 52} 3620 (1980).
Note the $g_3$-term in this reference is different from ours.

\bibitem{chain-mf}
D.J. Scalapino, Y. Imry and P. Pincus, Phys. Rev. B {\bf 11}, 2042 (1975).

\bibitem{jun}
Jun Zang, A. R. Bishop,
and D. Schmeltzer, Phys. Rev. B {\bf 52}, 6723 (1995); {\bf 54}, 9556 (1996).

\bibitem{cross}
M. C. Cross and D. S. Fisher, Phys. Rev. B {\bf 19}, 402
(1979).


\bibitem{gros} A. Fledderjohann and C. Gros, cond-mat/9612013.

\bibitem{haldane} F. D. M. Haldane, Phys. Rev. B {\bf 25}, 4925 
(Errata {\bf 26} 5257) (1982).

\bibitem{exact} 
H. Bergknoff and H.B. Thacker, Phys. Rev. D {\bf 19}, 3666 (1979).
 
\bibitem{note-log} There is generally a correction of
 $\log\delta$ due to the presence of frustration $\alpha_0$.
 See R. Chitrar et al., Phys. Rev. B {\bf 52} 6581 (1995).
 
 
 \bibitem{neel}
 S.B. Oseroff {\em et.al.}, Phys. Rev. Lett. {\bf 74}, 1450(1995).
 .
 \bibitem{neu-coex} L.P. Regnault {\em et al.}, 
  Europhys. Lett. {\bf 32}, 579 (1995);
 M. Poirier {\em et.al.}, Phys. Rev. B {\bf 52}, R6971(1995).

\bibitem{sol-la}
B. Horovitz, Phys. Rev. Lett. {\bf 46}, 742 (1981);
A.I. Buzdin, M.L. Kulic, and V.V. Tugushev, 
Solid State Commun. {\bf 48}, 483 (1983);
M. Fujita and K. Machida, J. Phys. Soc. Jpn. {\bf 53}, 4395 (1984).

\bibitem{book-sn} I.S. Gradshteyn and I.M. Ryzhik, {\it Tables of
Integrals, Series, and Products} (Academic Press, New York, 1965). 

\bibitem{note-sn}
Note that our expressions of soliton lattice solutions 
for $M(x)$ and $N(x)$ are
slightly different from those of Fujita and Machida\cite{sol-la}, but the
nature of these solutions are the same as shown in Fig.~(\ref{fig}).

\bibitem{note-exp} The relation between $\langle M(x)\rangle/\sqrt{\langle
N^2(x)\rangle}$ and $\tilde{Z}/Z$ is subtle, 
since $\langle S^z_{i,j+\mu}\rangle/
\langle S^z_{i,j}\rangle$ oscillates between $1-\epsilon\sim 1+\epsilon$.

\end{thebibliography}
\end{document}